\documentclass{osa-article}

\journal{osac}
\articletype{Research Article}

\usepackage{ulem} 
\usepackage[noabbrev,capitalise]{cleveref} 
\crefname{subsection}{subsection}{subsections}

\begin{document}

\setcopyright{}

\title{Image reconstruction through a multimode fiber with a simple neural network architecture}

\author{
  Changyan Zhu\authormark{1,4},
  Eng Aik Chan\authormark{2,4},
  You Wang\authormark{1},
  Weina Peng\authormark{3},
  Ruixiang Guo\authormark{2},
  Baile Zhang\authormark{1,2,5},
  Cesare Soci\authormark{1,2,6},
  and Yidong Chong\authormark{1,2,7}}

\address{\authormark{1}Division of Physics and Applied Physics, School of Physical and Mathematical Sciences, Nanyang Technological University, Singapore 637371, Singapore.}
\address{\authormark{2}Centre for Disruptive Photonic Technologies, Nanyang Technological University, 637371, Singapore, Singapore}
\address{\authormark{3}State Key Laboratory of Quantum Optics and Quantum Optics Devices, Institute of Opto-Electronics, Shanxi University, Taiyuan 030006, China}
\address{\authormark{4}These authors contributed equally to this work.}

\address{\authormark{5}blzhang@ntu.edu.sg}
\address{\authormark{6}csoci@ntu.edu.sg}
\address{\authormark{7}yidong@ntu.edu.sg}


\colorlet{b}{blue}
\colorlet{r}{red}
\colorlet{m}{magenta}
\colorlet{k}{black}

\newbool{showc}
\setbool{showc}{true}

\newcommand*{\ADel}[1]
	{
	\textcolor{r}{\sout{#1}} 
	}

\newcommand*{\textA}[1]
	{
	\ifbool{showc}
		{\textcolor{b}{#1}}{{\textcolor{k}{#1}}}
	}

\newcommand*{\commA}[1]
	{
	\textcolor{m}{\it{#1} } 
	}

\newcommand{\Figwt}{0.86}
\newcommand{\FigwtS}{0.54}

\begin{abstract}
Multimode fibers (MMFs) have the potential to carry complex images for endoscopy and related applications, but decoding the complex speckle patterns produced by mode-mixing and modal dispersion in MMFs is a serious challenge.  Several groups have recently shown that convolutional neural networks (CNNs) can be trained to perform high-fidelity MMF image reconstruction.  We find that a considerably simpler neural network architecture, the single hidden layer dense neural network, performs at least as well as previously-used CNNs in terms of image reconstruction fidelity, and is superior in terms of training time and computing resources required.  The trained networks can accurately reconstruct MMF images collected over a week after the cessation of the training set, with the dense network performing as well as the CNN over the entire period.
\end{abstract}

\section{Introduction}
  
Optical fibers have proven to be extremely useful for endoscopy and related applications \cite{fiberendoEric,fiberendoLee}.  Present commercial methods for transmitting images through fibers are based on single-mode fiber bundles \cite{porat2016fiberbundle,Shinde2017fiberbundle}, consisting of thousands of fibers each transmitting a single pixel.  It would be advantageous to instead transmit images in multimode fibers (MMFs), which are easy to fabricate and thinner than single-mode fiber bundles, and could potentially carry much more information.  However, there is a serious drawback: due to mode-mixing and modal dispersion, any image coupled into a MMF is transformed into a complex speckle pattern at the output \cite{Gover76}.  Researchers have devised various methods for reconstructing the input images from the speckle patterns, based on finding the complex transmission matrix of the MMF \cite{choi2012scanner, caravaca2013real, gu2015design, loterie2015digital} or phase retrieval algorithms \cite{popoff2010measuring, n2018controlled, n2018mode}.  However, such methods generally require extra apparatus for measuring the optical phase, or have difficulty scaling to large image sizes.

Another promising approach is to use a training set of {\it a priori} known inputs to teach an artificial neural network (NN) how to map MMF output images to input images.  This would not require additional interferometric equipment, and can potentially scale up to large image sizes.  The idea was proposed and investigated decades ago \cite{aisawa1991remote, matsumoto1990proposal, marusarz2001neural}, but only in recent years has it been shown to perform well for reconstructing images of reasonable complexity
\cite{borhani2018learning, rahmani2018multimode, fan2019deep, yang2019deep, rahmani2019multimode}, aided by improvements in computational power and NN software.  

These recent advances in NN-aided MMF image reconstruction have focused on deep convolutional neural networks 
(CNNs) \cite{borhani2018learning, rahmani2018multimode, Kakkava2019, fan2019deep, yang2019deep, rahmani2019multimode}.  Unlike 
traditional dense NNs \cite{braspenning1995artificial}, CNNs use convolution operations instead of general matrix multiplication 
within the NN layers \cite{Goodfellow-et-al-2016}, inspired by biological processes in visual perception.  CNNs have enjoyed immense 
recent success in computer vision \cite{Rawat2017}, making it natural to investigate using them for MMF image reconstruction.  
They have also been applied to the related problem of image reconstruction in scattering 
media \cite{li2018deep,li2018imaging,Turpin2018Control,sun2019image}.  However, there are grounds to question 
how well-suited CNNs are for analyzing speckle patterns such as those produced by MMFs, which are very different 
from the natural images commonly dealt with in computer vision.  In MMF images, information is encoded not just locally 
but in the global distribution of speckles \cite{Redding2012, Pauwels2019, Kakkava2019}, whereas the localized receptive 
fields in convolutional layers are designed to extract relevant local features (such as edges) in natural images, rather 
than long-range spatial structures \cite{Linsley2018}. Traditional dense NNs  can extract information from both local and global features due to the presence of the fully-connected hidden layers.

This paper investigates the performance of dense NNs and CNNs for MMF image reconstruction. Whereas the earliest papers on NN-aided MMF 
image reconstruction used dense NNs \cite{aisawa1991remote, matsumoto1990proposal, marusarz2001neural}, most recent studies have concentrated 
on using CNNs \cite{borhani2018learning, rahmani2018multimode, Kakkava2019, fan2019deep, yang2019deep, rahmani2019multimode}.  (One exception to 
this trend was the study by Turpin \textit{et al.}~of both dense NNs and CNNs for transmission control in scattering media and 
MMFs~\cite{Turpin2018Control}.) To our knowledge, there has been no direct comparison between the two NN architectures in the context 
of MMF image reconstruction. The recent popularity of CNNs for this task is predicated on the local feature extraction capability of CNNs being useful for descrambling MMF images. Our comparison of dense NNs and CNNs should be useful to the community in testing this assumption.

Our principal comparison is between (i) the single hidden layer dense neural network (SHL-DNN), one of the simplest 
dense NN architectures, and (ii) U-Net, a CNN originally developed for biomedical imaging \cite{ronneberger2015u}, 
which has recently been used for MMF image reconstruction \cite{rahmani2018multimode}.  We do not compare very deep 
CNNs such as Resnet \cite{rahmani2018multimode} or  generative adversarial networks \cite{ZhangGAN}, as these require 
much greater computational resources and longer training times, and thus seem ill-suited to the MMF image 
reconstruction problem. After optimizing both types of NNs (SHL-DNN and U-Net), we find that the SHL-DNN 
achieves a similarly high reconstructed image fidelity with shorter training time and less network complexity 
than the U-Net. For one of our reference datasets, SHL-DNN achieves a saturation Structural Similarity Index 
Measure (SSIM) \cite{wang2004image} of $0.775$ in $16$ minutes and the U-Net achieves a saturation SSIM of $0.767$ 
in $3.5$ hours on the same computer.  The SHL-DNN has $20$ million trainable parameters, and the U-net has $31$ 
million trainable parameters.  We also validated both NNs using images collected up to $235$ hours 
after the images in the training set; both NNs continue to perform well in image reconstruction.  Moreover, we 
tested a ``VGG-type'' NN, which combines convolutional and dense layers, and found that it offers no additional 
performance advantage over the SHL-DNN.

\section{Experimental Setup}

\subsection{Multimode fiber imaging}

\begin{figure}
\centering
\includegraphics[width=\Figwt\linewidth]{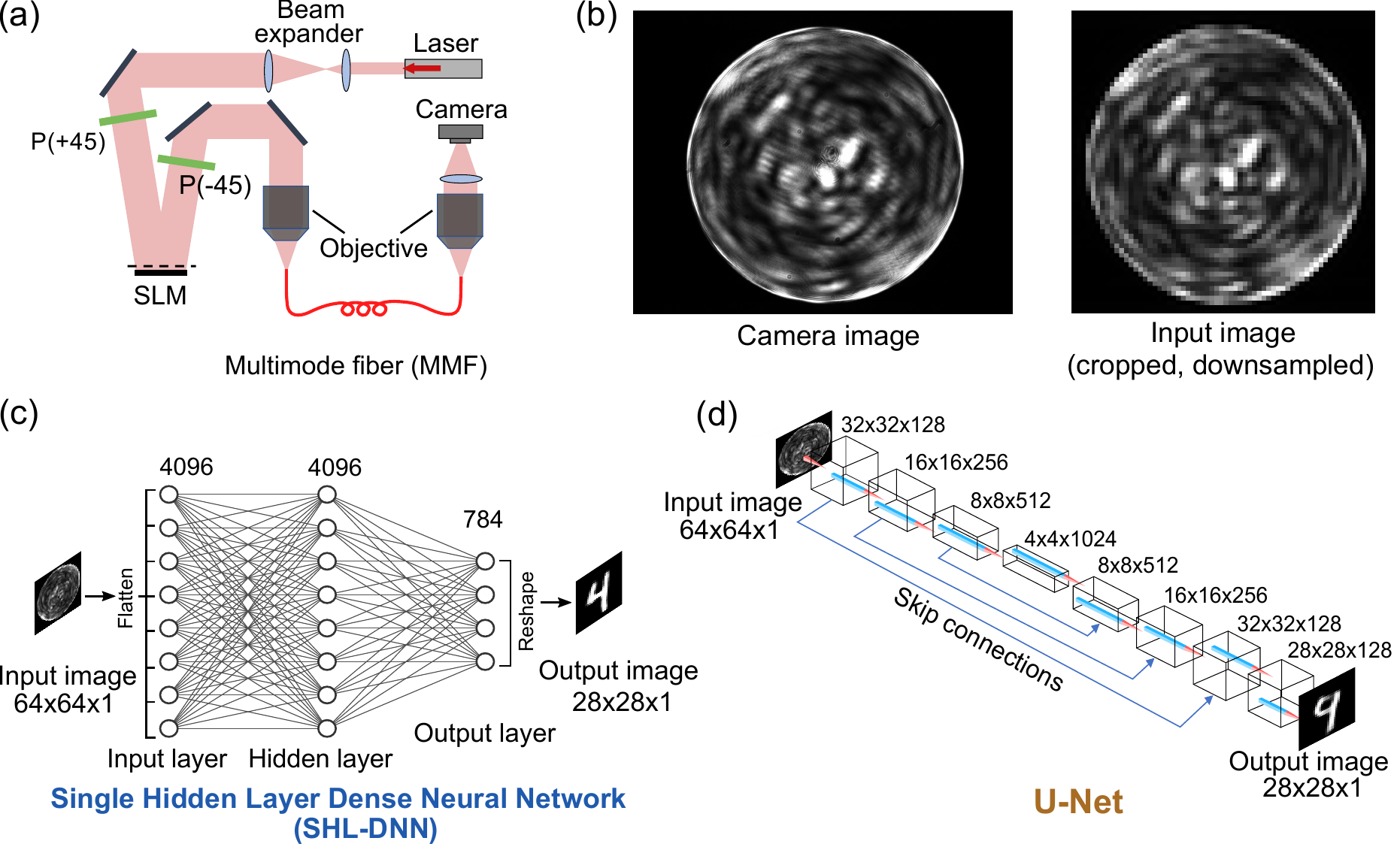}
\caption{(a) Experimental setup for imaging through a multimode fiber. A laser beam is expanded and reflected off a spatial light modulator (SLM), which together with a pair of polarizers (P) generates an intensity modulated image.  The beam is coupled into a multimode fiber (MMF), and the distal end is imaged by a camera. (b) Example of a scrambled image from the MMF.  The ground truth image is a digit from the MNIST database (see Fig.~\ref{fig:fig2}).  The left panel shows the full-resolution ($1280\times1080$ pixels) camera image; the right panel shows the  cropped and downsampled $64\times64$ image fed to the neural network.  (c) Schematic of a single hidden layer dense neural network (SHL-DNN) with $4096$ nodes in the hidden layer.  The input image is flattened at the input layer, and the output is reshaped into a two-dimensional image. (d) Schematic of a U-Net consisting of contracting convolutional layers, an intermediary layer, and expanding convolutional layers.  For each convolutional layer, the size $a\!\times\!b\!\times\!c$ refers to $a\!\times\!b$ pixels with $c$ filters (image depth).  Skip connections concatenate the outputs from successive contracting layers with the corresponding expanding layers.}
\label{fig:fig1}
\end{figure}

The optical setup is shown in Fig.~\ref{fig:fig1}(a).  A collimated beam from a diode laser with an operating wavelength of 808 nm (Thorlabs LP808-SF30) is expanded and directed onto a spatial light modulator (SLM) (Hamamatsu X13138-02). Along with two orthogonal polarizers, the SLM generates a programmable spatial modulation in the intensity of the light beam.

The modulated beam is coupled into a one meter long multimode fiber (MMF) (Thorlabs FT400EMT) via a matching collimator (NA $0.39$).  The distal end of the MMF is imaged with a CMOS camera (Thorlabs DC1545M).  The camera images consist of complicated speckle patterns, as shown in the left panel of Fig.~\ref{fig:fig1}(b), with no apparent relation to the ground truth images from the SLM.  The camera images have $1280\times1080$ pixel resolution; to obtain a tractable dataset, we crop and downsample to $64\times64$ using the Lanczos algorithm \cite{LanczosResampling}, as shown in the right panel of Fig.~\ref{fig:fig1}(b).

By operating the SLM with a refresh rate of 0.9 Hz (which allows for the generation of stable and distortion-free images), we accumulate one dataset of $61524$ MMF images collected over approximately $19$ hours for training and several datasets that spans across $235$ hours. The ground truth images are drawn equally from (i) the MNIST digit dataset containing handwritten digits in various styles \cite{lecun2010mnist}, and (ii) the MNIST-Fashion dataset containing images of clothing and apparel \cite{xiao2017fashionmnist}.  The MNIST digit dataset is used for most of the experiment; the MNIST-Fashion dataset is used in  Section~ \ref{sec:transfer}.

The MNIST and MNIST-Fashion ground truth images are $28\times28$, whereas the MMF-derived images in the dataset are $64\times64$.  Conceptually, there is no reason to restrict the MMF images (NN inputs) to the same size as the ground truth images (and NN outputs), as was the practice in earlier studies \cite{borhani2018learning, rahmani2018multimode}.  Intuitively, higher resolutions for the MMF images should be advantageous, as the image reconstruction algorithm is given more information to work with, subject to the constraints of trainability and computer memory capacity.  The effects of varying the input size are studied in Section~\ref{sec:reconstruction}.  Both the ground truth and MMF images have 8 bits of dynamic range.

\subsection{Neural networks} \label{sec:NN}

We mainly investigate and compare two NN architectures for efficacy in MMF image reconstruction: a single hidden layer dense neural network (SHL-DNN) and the convolutional neural network U-Net.  (A third architecture, a hybrid convolutional/dense network, is briefly discussed in Section~\ref{sec:vgg}.)

Dense NNs are the most elementary architecture for NN-based machine learning.  The earliest papers on NN-aided MMF image reconstruction utilized dense NNs \cite{aisawa1991remote, matsumoto1990proposal, marusarz2001neural}, but were constrained by the lower levels of computational power then available. We implement the SHL-DNN shown in Fig.~\ref{fig:fig1}(c), featuring a hidden layer of $4096$ nodes sandwiched between input and output layers, with dense interlayer node connections.  Each $64\times64$ input image is flattened and inserted into the input layer, which has $64^2 = 4096$ nodes.  The hidden layer and output layer have sigmoid activation functions.  The result from the output layer (which has $28^2 = 784$ nodes) is reshaped into a $28\times28$ image that can be compared to the ground truth image.

Convolutional neural networks (CNNs) have been applied to the MMF image reconstruction problem by several recent 
authors \cite{borhani2018learning, rahmani2018multimode, Kakkava2019, fan2019deep, yang2019deep, rahmani2019multimode}.  
Here, we employ the U-Net architecture, which Rahmani \textit{et al.}~have previously used for MMF image reconstruction 
with the MNIST digit dataset \cite{rahmani2018multimode}.  As shown in Fig.~\ref{fig:fig1}(d), the input is $64\times 64$ and the output is $28\times28$, 
the same as for the SHL-DNN.  The network consists of a sequence of convolutional and pooling layers leading to a $4\times4\times1024$ 
intermediary layer, followed by a sequence of convolutional upsampling layers. Batch normalization is applied after each convolutional 
layer. Each convolutional layer has a ReLU activation function, and the output layer has a sigmoid activation function (similar to the SHL-DNN). We follow the typical U-Net architecture design 
rule \cite{ronneberger2015u} wherein a halving of the layer dimensions is accompanied by a doubling of the number of filters 
(image depth), and vice versa. There are also auxiliary skip connections that aid image localization \cite{ronneberger2015u}.

The U-Net architecture contains numerous hyperparameters such as the number of layers, convolutional filter depths, batch size, etc. We tested the effects of varying these hyperparameters, and the ``baseline'' configuration shown in Fig.~\ref{fig:fig1}(d) gives
the best results.  Notably, in this configuration the filter depths are four times what was used in Rahmani \textit{et al.}~\cite{rahmani2018multimode}.

Each NN is trained using Adam optimization with a batch size of $256$ images, and an early stopping condition of $100$ epochs after validation losses stop improving. We find that batch-normalization regularization is crucial for the U-Net to perform well, but dropout regularization is better for the SHL-DNN.  Little performance improvement is observed when the batch size adjusted between $128$ and $1024$; a much larger 
batch size ($27685$) drastically lengthens training times.  For the objective function, the NN output is compared against the ground truth (MNIST digit or MNIST-Fashion) image via the Structural Similarity Index Measure (SSIM) \cite{wang2004image}, a well-established metric for quantifying the similarity between structured 
images (see Supplementary Materials).  All training was performed on the same computer (Intel Xeon Gold $5218$ with NVIDIA Quadro RTX $5000$ GPU).

\section{Results} \label{sec:Results}

\subsection{Image reconstruction fidelity}
\label{sec:reconstruction}

\begin{figure}
\centering
\includegraphics[width=\Figwt\linewidth]{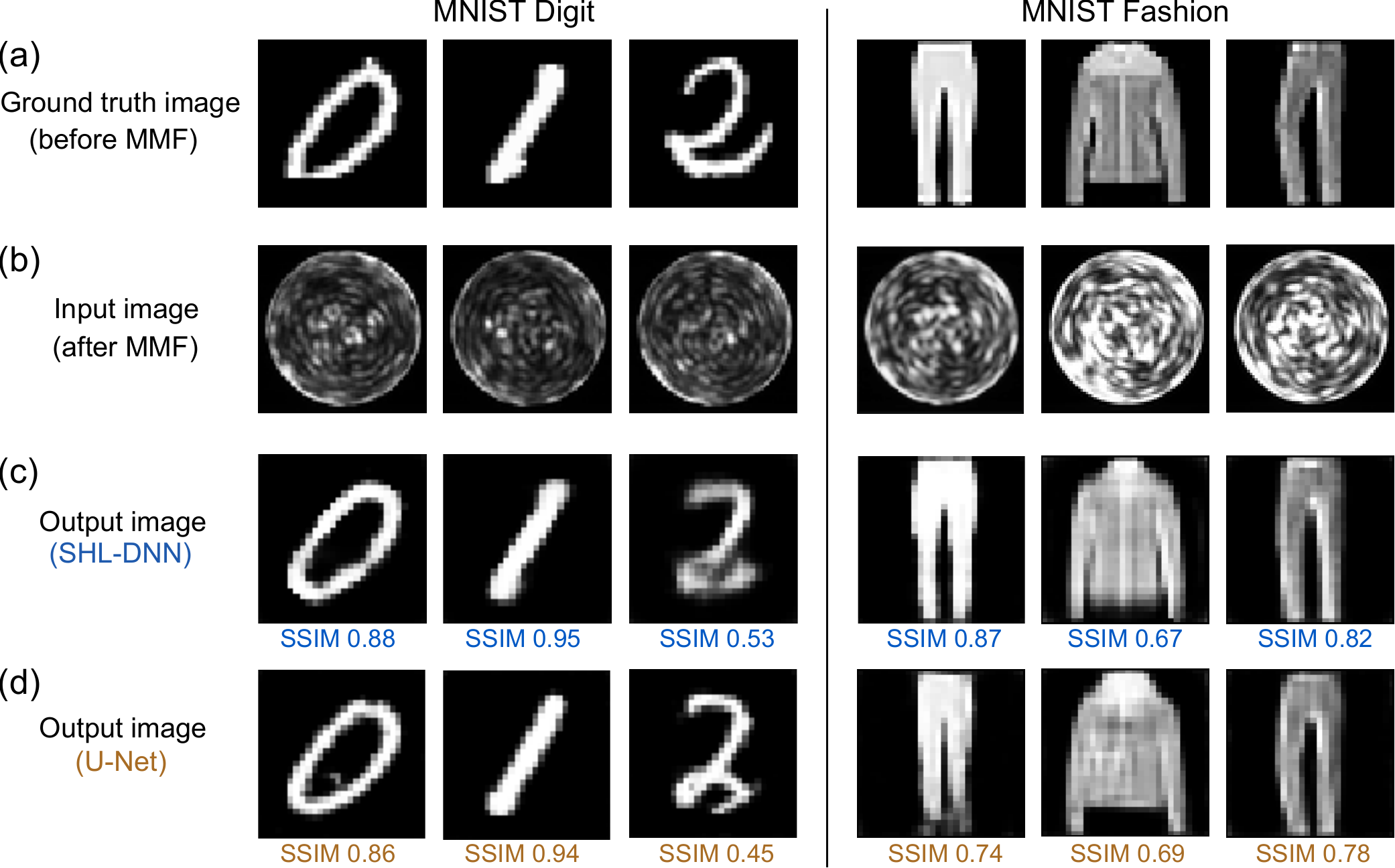}
\caption{Demonstration of MMF image reconstruction on the MNIST digit dataset (first three columns) and the MNIST-Fashion dataset (last three columns).  
(a) A representative sample of $28\times 28$ ground truth images.  
(b) The corresponding $64\times64$ images obtained from the MMF.  
(c) Reconstructed $28\times28$ images produced by the SHL-DNN.  The structural similarity (SSIM) relative to the ground truth 
image is shown below each reconstructed image.  (d) The corresponding results produced by the U-Net.}
\label{fig:fig2}
\end{figure}

We train the SHL-DNN and U-Net using $30762$ MMF images from the first 19 hours of the data collection run.  The ground truth images drawn from either the MNIST digit dataset \cite{lecun2010mnist} or the MNIST-Fashion dataset \cite{xiao2017fashionmnist}; separate instances of each network are trained for the two respective datasets. In each case, we assign $27685$ images for training and the remaining $3077$ for validation.  The training and validation images are initially drawn randomly from across the collection period (the role of collection time will be investigated later, in Section~\ref{sec:stability}).

Fig.~\ref{fig:fig2} shows the results of MMF image reconstruction for six representative images from the validation set, three from the MNIST digit datasets and three from the MNIST-Fashion sets.  The fully-trained SHL-DNN and U-Net both recover the ground truth images with remarkable fidelity [Fig.~\ref{fig:fig2}(a) and (c)--(d)], despite the lack of human-discernable patterns in the MMF images [Fig.~\ref{fig:fig2}(b)].  The SHL-DNN and U-Net achieve similar fidelity, as corroborated by the similar SSIM scores, for both types of images.

\begin{figure}
\centering
\includegraphics[width=\Figwt\linewidth]{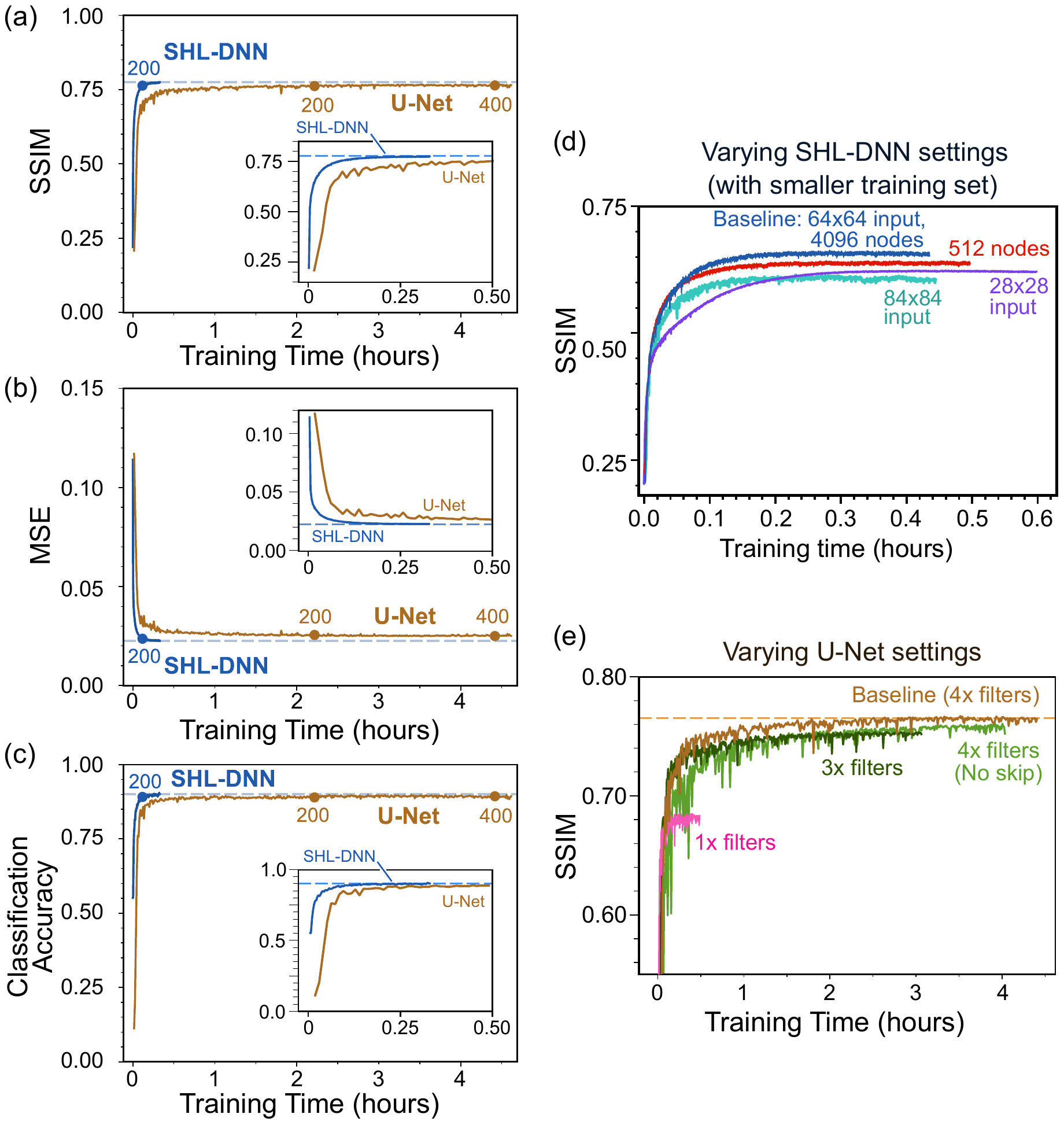}
\caption{(a)--(c) Training curves for SHL-DNN and U-Net, using SSIM as the objective function and with 27685 training images and 3077 validation images.  Epoch numbers are indicated by the numbered circles on each curve.  (a) SSIM versus training time.  (b) Mean squared error (MSE) versus training time. (c) Classification accuracy versus training time, obtained by feeding the output images from each neural network into an auxiliary classifier network, serving as a measure of legibility.  In (a)--(c), the converged performance measure for the SHL-DNN is indicated by horizontal dashes.  (d) SHL-DNN performance with different settings, calculated with a training set of $8709$ images: the baseline network used in (a)--(c) and depicted in Fig.~\ref{fig:fig1}(c), with $64\times64$ inputs and $4096$ hidden layer nodes (blue), a network with $28\times28$ inputs (purple), a network with $84\times84$ inputs (green), and a network with 512 hidden layer nodes and $64\times64$ inputs (red).  (e) U-Net performance for different settings: the baseline U-Net with ``$4\times$'' filters [used in (a)--(c) and depicted in Fig.~\ref{fig:fig1}(d)] (brown), removed skipping layers (light green), ``$3\times$'' filters (dark green), and ``$1\times$'' filters (pink).  In the last case, convergence is achieved about as quickly as the SHL-DNN, but at significantly lower SSIM.}
\label{fig:fig3}
\end{figure}

Fig.~\ref{fig:fig3}(a) shows the SHL-DNN and U-Net training curves for the MNIST digit dataset (the results for MNIST-Fashion are similar; see Supplementary Materials).  We plot the training curves against elapsed time to allow for fairer comparisons, since the two networks have very different training times per epoch.  The performance of the SHL-DNN saturates at SSIM $0.775$, comparable to SSIM $0.767$ for the U-Net.  Fig.~\ref{fig:fig3}(b) and (c), we compare the performance of both networks on two other common metrics: the mean squared error (MSE) and the resulting classification error for the validation set (however, the training still uses SSIM for the objective function).  The classification error is meant to characterize the overall legibility of the reconstructed digits, and is obtained by passing the NN outputs to an auxiliary digit classifier (\texttt{mnist\_cnn.py} from Keras \cite{keras}).  The results from these alternative measures are similar to what was obtained from the SSIM.  The SHL-DNN achieves MSE $2.25\times10^{-2}$ and classification accuracy $0.90$, while the U-Net achieves MSE $2.48\times10^{-2}$ and classification accuracy $0.90$.

Although the two networks yield similar image reconstruction fidelity, the SHL-DNN can be trained more quickly.  To reach its saturation SSIM (i.e., triggering of the stopping condition), the SHL-DNN takes $462$ epochs and $16$ minutes, whereas the U-Net takes $318$ epochs and $3.5$ hours.  The training time per epoch is $20$ times faster for the SHL-DNN.

We systematically investigated the effects of various NN settings, and found that no further major performance improvements are achievable without increasing the training set size.  (For these hyperparameter studies, a smaller training set of 8709 images was utilized.)  For the SHL-DNN, the choice of input image size appears to play an important role.  As shown in Fig.~\ref{fig:fig3}(d), for a smaller input image size ($28\times28$) the SSIM saturates at a lower value, which can be ascribed to the NN having less information available for image reconstruction.  But having inputs that are too large, such as $84\times84$, also leads to a lower SSIM compared to our baseline choice of $64\times64$.  The SHL-DNN performance decreases when the number of hidden layer nodes is reduced below the baseline value, as shown by the red curve in Fig.~\ref{fig:fig3}(d) for the 512 node case.  On the other hand, further increasing the number of hidden layer nodes increases the training time without significant improvement in the saturated SSIM (see Supplementary Materials).  Moreover, the number of hidden layer nodes seems to have negligible influence on the optimal input image size.

As for the U-Net, one setting that notably affects performance is the number of convolutional filters. We denote the number of U-Net filters used in Rahmani \textit{et al.}~\cite{rahmani2018multimode} as ``$1\times$''.  The saturated SSIM score increases as the number of filters in increased up to $4\times$, which is the baseline value that we adopted.  Further increases in the number of filters leads to a substantial increase in training time, without significant performance improvement. Another possible setting is the number of convolutional layers; we verified that deeper or shallower U-Net structures adversely affect the performance.  Moreover, we find that removing the skip connections leads to a slight decrease in performance slightly; hence, the skip connections are included in our baseline configuration (although the reason these connections are advantageous for the present task is somewhat unclear).  Some of these comparisons are shown in Fig.~\ref{fig:fig3}(e).

After these optimization studies, we arrive at the SHL-DNN and U-Net configurations shown in Fig.~\ref{fig:fig1}(c)--(d).  In these configurations, the SHL-DNN has $20$ million trainable parameters and takes 39.9 million FLOPs per forward pass, and the U-net has $31$ 
million trainable parameters and takes 62.8 million FLOPs per forward pass.

\subsection{Performance over time}
\label{sec:stability}

\begin{figure}
\centering
\includegraphics[width=\Figwt\linewidth]{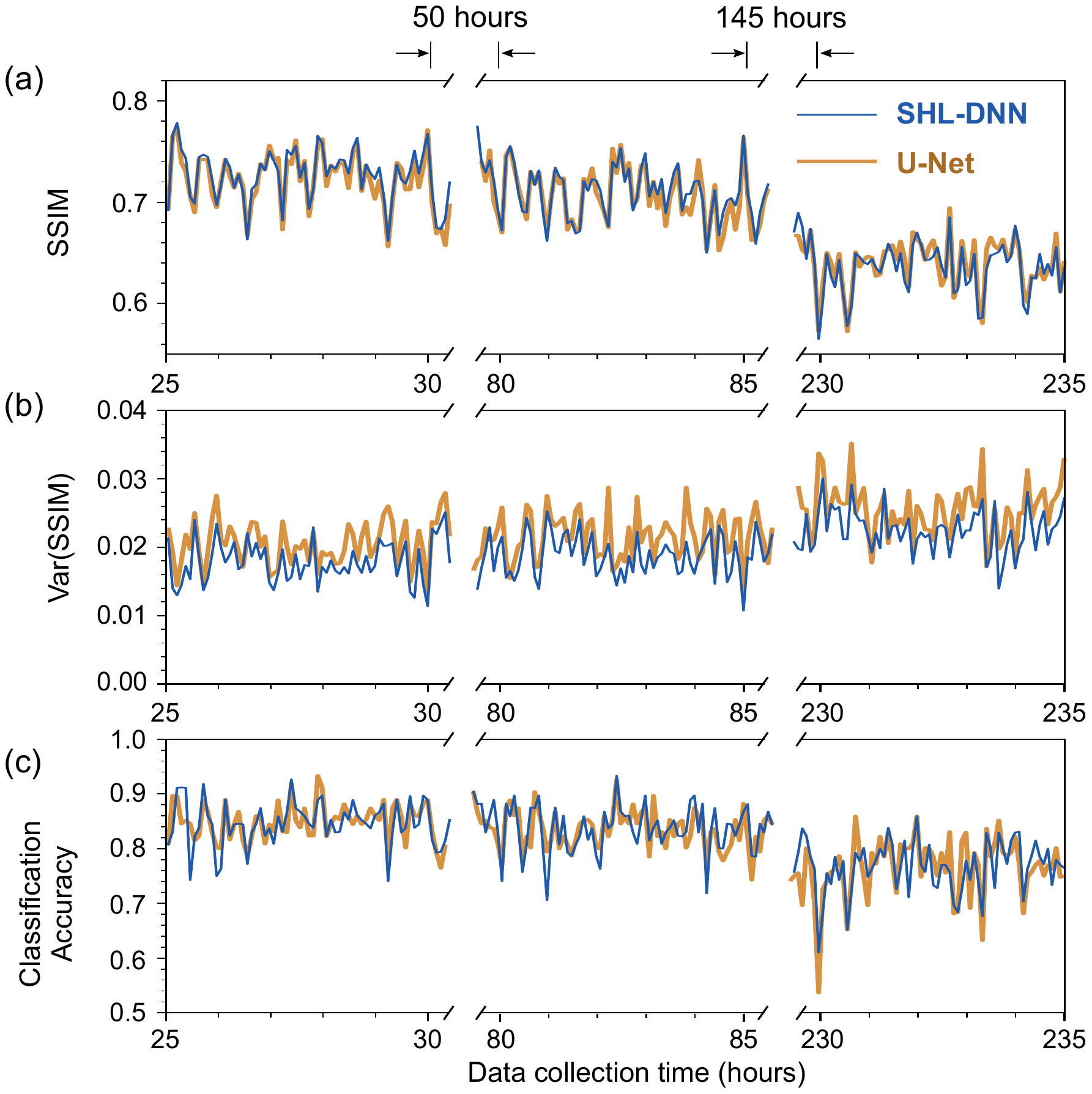}
\caption{MMF image reconstruction metrics using data collected at different times subsequent to the training set.  The SHL-DNN and U-Net are trained using $27685$ images collected over $19$ hours, and then validated against images collected over the subsequent $235$ hours.  The time axis is divided into $5$ minute bins with 137 validation images per bin.  (a) SSIM.  (b) Variance of SSIM, corresponding to the spread of the SSIM in each $5$ minute bin.  (c) Classification accuracy, obtained by feeding the output images from each neural network into an auxiliary high-accuracy classifier.}
\label{fig:fig4}
\end{figure}

It is interesting to ask whether the image reconstruction ability of the NNs is persistent, or whether it degrades over time due to a drift in the MMF's transmission characteristics.  Such temporal changes can be caused by thermal and mechanical perturbations of the environment, which induce minute deformations of the fiber.

To address this question, we validate the NNs (trained using images from the first $19$ hours of the dataset) against images collected during the subsequent $235$ hours.  The results are shown in Fig.~\ref{fig:fig4}.  The validation data are sorted by collection time and batched into $5$ minute intervals.

In terms of both SSIM and digit classification accuracy, the image reconstruction performance for both NNs fluctuates over time, but is overall remarkably robust.  It can be noticed in Fig.~\ref{fig:fig4}(a) and (c) that the performance fluctuations for the SHL-DNN and U-Net are correlated over time.  In fact, their SSIM scores have a correlation coefficient of $0.950$.  This implies that the performance fluctuations are caused by the MMF undergoing physical fluctuations in its transmission characteristics (relative to the training set), which simultaneously impacts the performance of both NNs.  Over the $235$ hour period, we observe only a slight long-term degradation in performance (both in terms of SSIM and digit classification accuracy), indicating that there is neligible sustained ``drift'' in the MMF's transmission characteristics.  Over the entire experimental period, the SHL-DNN and U-Net consistently have similar performance, with SSIM variance of about $0.02$.

\subsection{Hybrid neural network}
\label{sec:vgg}

\begin{figure}
\centering
\includegraphics[width=\Figwt\linewidth]{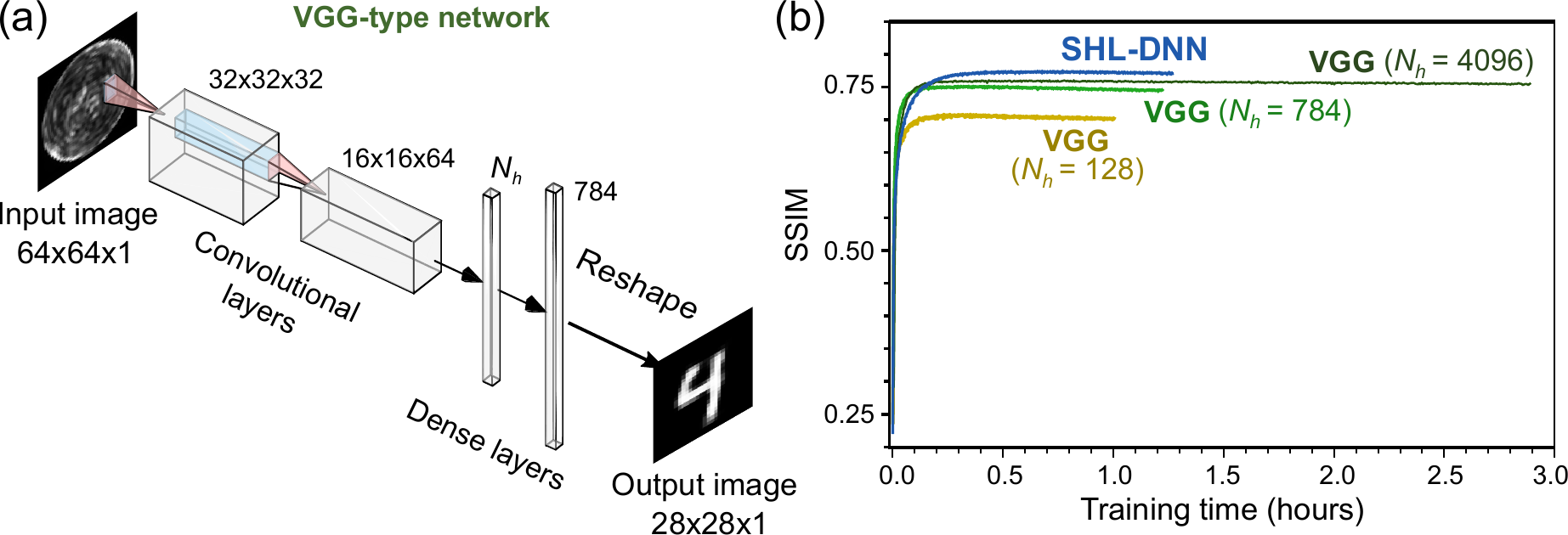}
\caption{Performance of a VGG-type network for MMF image reconstruction.  (a) Schematic of the VGG-type network, which consists of two convolutional layers, a dense hidden layer with $N_h$ nodes, and a dense output layer. (b) Training curves for SHL-DNN and VGG-type networks: the baseline SHL-DNN corresponding to Fig.~\ref{fig:fig1}(c), with 4096 hidden layer nodes (blue), and VGG-type networks with $N_h = 4096$ (dark green), $N_h = 784$ nodes (light green), and $N_h = 128$ (yellow).}
\label{fig:fig5}
\end{figure}

Rahmani \textit{et al.}~\cite{rahmani2018multimode} studied the use of another type of NN for unscrambling MMF images: a hybrid convolutional and dense network of the type pioneered by Oxford's Visual Geometry Group (VGG).  VGG-type networks are typically used for classification \cite{simonyan2014deep}, and they were used in Ref.~\cite{rahmani2018multimode} for digit classification with the MNIST digit dataset.  In this paper, we are mainly interested in image \textit{reconstruction} rather than \textit{classification}.  Nonetheless, it is helpful to study the performance of a VGG-type network for this purpose, as a further test of the usefulness of convolutional layers for extracting structural information from MMF images.

We implement a simple VGG-type network as shown in Fig.~\ref{fig:fig5}(a), consisting of two convolutional layers, a hidden dense layer with $N_h$ nodes, and a dense output layer.  Fig.~\ref{fig:fig5}(b) shows the training curves for VGG-type networks with several choices of $N_h$, as well as for the baseline SHL-DNN.  When $N_h$ is equal to the number of hidden layer nodes in the SHL-DNN, the saturated SSIM is $0.71$---comparable to but certainly not better than the SHL-DNN (SSIM $0.775$).  For smaller values of $N_h$, the performance is substantially worse.  We also investigated reversing the configuration by placing the dense layers at the input and the convolutional layers at the output, but this not produce any improvement.  These results seem to bolster the case that convolutional input layers do not provide additional benefits for MMF image reconstruction, a point that will be further discussed in Section~\ref{sec:discussion}.

\subsection{Transfer learning and alternate image set}
\label{sec:transfer}

\begin{figure}
\centering
\includegraphics[width=\Figwt\linewidth]{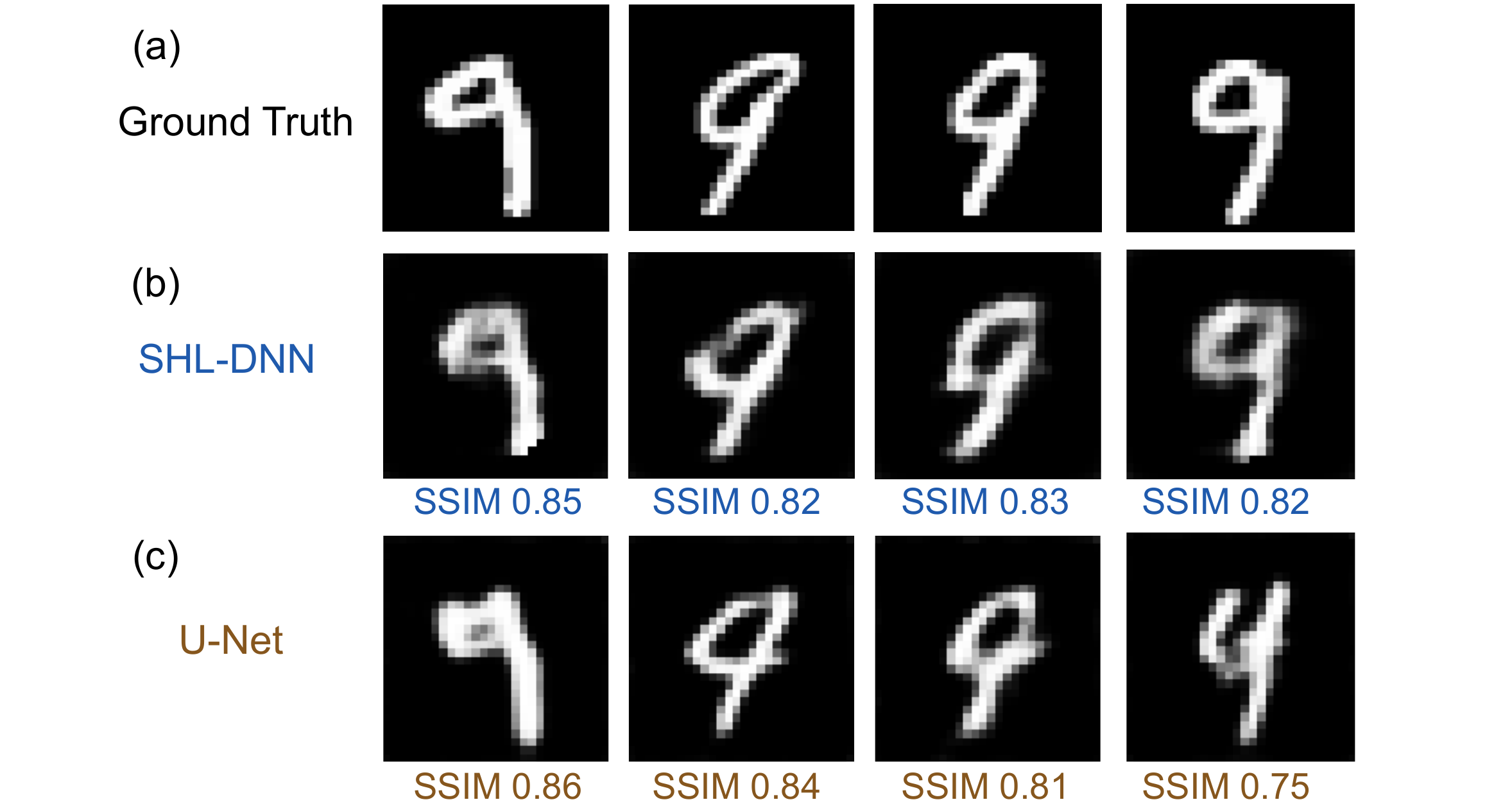}
\caption{Reconstruction of images of the digit 9 from the MNIST digit dataset, using a SHL-DNN trained with a 
modified MNIST digit dataset excluding all instances of the digit 9.  
(a) Ground truth images.
(b) Reconstructed images of digit '9' by the SHL-DNN trained with the modified dataset.  
(c) Reconstructed images of digit '9' by the U-net trained with the modified dataset.  
SSIM scores are shown below the reconstructed images. }
\label{fig:fig6}
\end{figure}

Transferability is a common concern in machine learning.  In the present context, one may ask whether NNs trained using one kind of ground truth image---say, MNIST digits---can successfully reconstruct more general images.  In other words, are the networks broadly capable of undoing the effects of mode mixing in the MMF, or are they merely recognizing patterns that are highly specific to the sort of images in the training set?

To investigate this, we train the SHL-DNN by withholding one digit from the MNIST digit dataset, and validating it against the omitted digit.  
Fig.~\ref{fig:fig6}(a) shows representative results for the case of an omitted digit `9'.  Although this SHL-DNN has not seen any examples 
based on the digit `9', it reconstructs the images reasonably well, albeit with lower SSIM.  Here, the training set (with `9' excluded) 
has 14565 images, and the other network settings are the same as in the baseline network described in Section~\ref{sec:reconstruction}.
Over 1000 instances of the digit `9', the mean SSIM is $0.72$, compared to SSIM $0.86$ for a validation set of $2913$ images that exclude 
the digit `9'. The performance of U-net is quite similar to SHL-DNN: the mean SSIM is $0.70$ over 1000 instances of the digit '9'.


When we attempt to reconstruct MNIST-Fashion images using a SHL-DNN trained on MNIST digits, or vice versa, the results are extremely 
poor (SSIM close to zero).  Likewise, when we attempt to reconstruct images consisting of random uncorrelated pixel intensities, 
all three trained networks (SHL-DNN, U-Net, and VGG-type) give very poor results; over $1000$ images, the MSE is in the range 
of $0.08-0.09$ for all the three networks, comparable to the nascent training stage of Fig.~\ref{fig:fig3}(b).

\section{Discussion}
\label{sec:discussion}

We find that CNNs offer no performance advantage over the traditional dense NN architecture for MMF image reconstruction.  
In fact, the tested SHL-DNN is able to produce the same results as U-Net with much shorter training time and less network complexity, 
and this seems to be robust over various different NN settings. Moreover, a VGG-type hybrid convolutional/dense NN offers no obvious improvement over the SHL-DNN. These results suggest that the SHL-DNN, though a simple architecture, already gives a ceiling for the performance of neural networks on MMF image reconstruction.
For practical real-time imaging applications, simpler NN architectures may be desirable as they can be trained more quickly and with 
fewer computational resources.

One interpretation of the situation is that convolutional layers, though well-suited to extracting local features in natural images, do not provide any special advantage in processing speckle patterns of the sort produced by MMFs \cite{Redding2012, Pauwels2019}.  It would be interesting to explore modifications to the CNN scheme, or preprocessing schemes for the speckle pattern, to improve performance \cite{Kakkava2019}.

The trained NNs can reliably reconstruct images collected hours after the training set; we observe short-term performance fluctuations that can be ascribed to environmental effects, but no degradation corresponding to a long-term drift in the fiber transmission characteristics.  The main bottleneck in terms of training is the relatively low refresh rate of the spatial light modulator.

The NNs perform poorly on images that are too different from those in the training set, which is a common problem with NN-based machine learning.  Recently, Caramazza \textit{et al.}~have demonstrated using an optimization algorithm to learn the complex transmission matrix for MMF image reconstruction \cite{caramazza2019transmission}, which bypasses the transfer learning limitations of the NN approach.  However, this method requires much more computer memory, and the resulting image fidelity is lower; from our testing based on MNIST digits, the SSIM scores are in the range 0.2 -- 0.5, compared to $\sim 0.75$ for the SHL-DNN.  In the future, it would be interesting to attempt to combine these two approaches in a way that overcomes their individual limitations.

\setlength{\parskip}{2em}

\textbf{Funding}. The authors acknowledge support from the Singapore Ministry of Education Tier 3 Grant MOE2016-T3-1-006 and Tier 1 Grant RG187/18.

\textbf{Disclosures}. The authors declare no conflicts of interest.


\section{Supplementary Materials}

\appendix
\section{SSIM and MSE}

The SSIM (Structural Similarity Index) is a commonly-used perceptual metric based on visible structures in the image, developed by Wang \textit{et al.}~\cite{wang2004image}.  It is defined as
\begin{equation}
\textrm{SSIM}(x,y) = l(x,y)^{\alpha} \cdot c(x,y)^{\beta} \cdot (x,y)^{\gamma}
\end{equation}
where
\begin{equation}
l(x,y)^{\alpha} = \frac{2 \mu_x \mu_y + c_1}{\mu_x^2 + \mu_y^2 + c_1 }, \;\;
c(x,y) =\frac{ 2 \sigma_x \sigma_y + c_2}{\sigma_x^2 + \sigma_y^2 + c_2}, \;\;
s(x,y) = \frac{\sigma_{xy}+c_3}{\sigma_x \sigma_y + c_3}
\end{equation}
are measures for luminance, contrast, and structural similarity respectively.  Here, $x$ and $y$ denote two images to be compared, $\mu_x$ and $\mu_y$ are their average values, $\sigma_x$ and $\sigma_y$ are the standard deviation, and $\sigma_{xy}$ is the covariance.   We set
$c_1= (k_1 L)^2$, $c_2= (k_2 L)^2$, and $c_3 = c_2/2$. $k_1= 0.01$ and $k_2 = 0.03$. $L = 1$ is the dynamic range of the pixel-values. $\alpha$, $\beta$, and $\gamma$ are the weights for each feature, which we set to 1.

When training the neural networks (NNs), we use the SSIM to define a loss function according to
\begin{equation}
Loss_{\textrm{SSIM}} = 1 - \textrm{SSIM}.
\end{equation}

Another measure of the similarity between two images $X$ and $Y$, the Mean Squared Error (MSE), is defined as
\begin{equation}
  MSE = \frac{1}{MN} \sum_1^M \sum_1^N [X(m,n)-Y(m,n)]^2
\end{equation}
As shown in Fig.~3 of the main text, the SSIM and MSE produce similar results when used to evaluate NN performance.

\section{Neural network settings}

To optimize the performance of each NN, we tested several different hyperparameter choices.

\subsection{SHL-DNN versus multilayer perceptron}

In several applications of dense neural networks (DNNs), it is advantageous to increase the depth of the NN since this 
increases the representational capacity of the network.  In the context of multimode fiber (MMF) image reconstruction, 
we investigated the performance of dense networks with different depths.

Fig.~7(a) shows the training curves for two dense NNs with different numbers of hidden layers, but the same total number 
of trainable parameters.  The SHL-DNN has 4096 nodes in the hidden layer, and the 3-layer DNN contains 2164 nodes per hidden 
layer.  We use the same early-stopping condition (100 epochs after validation losses stop improving).  The SHL-DNN saturates 
at SSIM of $0.775$, significantly higher than the 3-layer DNN, which has SSIM of $0.766$.  Hence, a multilayer structure does 
not appear to be advantageous for image reconstruction fidelity.

\begin{figure}
\centering
\includegraphics[width=\textwidth]{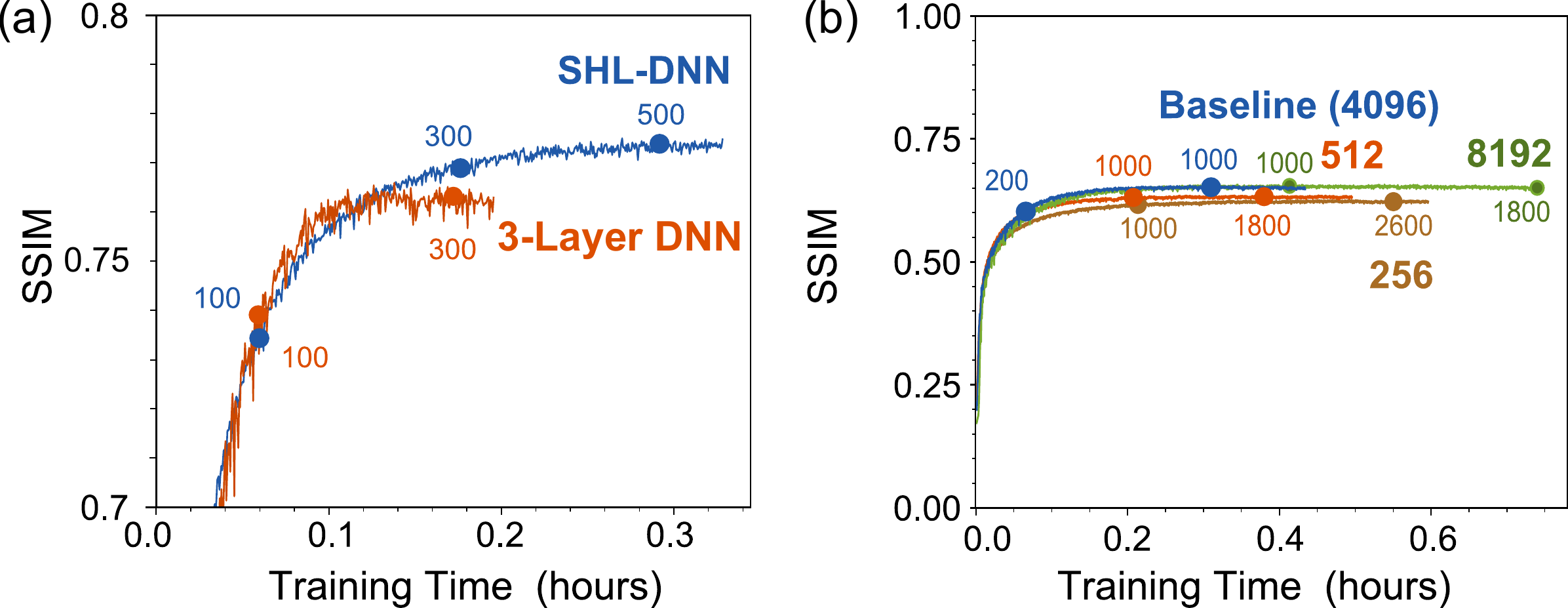}
\caption{(a) Training curves for dense neural networks with a single hidden layer (blue) and 3 hidden layers (orange).  The networks have the same total number of trainable parameters.  (b) Training curves for SHL-DNNs with different numbers of hidden nodes: 256 (brown), 512 (orange), 4096 (baseline, blue), and 8192 (green).}
\label{fig:figs1}
\end{figure}

\subsection{Number of hidden layer nodes in SHL-DNN}

In Fig.~7(b), we plot the training curves SHLs with different numbers of hidden layer nodes ($256$, $512$, $4096$, and $8192$).  
It can be seen that negligible improvement is achieved by going from $4096$ to $8192$ nodes.

\subsection{Activation units in SHL-DNN}

Fig.~8(a) shows the training curves for SHL-DNNs with different activation units in the hidden layer: tanh, ReLU, and sigmoid.  The sigmoid activation appears to perform the best, and this was the one we used for the SHL-DNNs in the main text.

Also shown here is the training curve for a SHL-DNN without a sigmoid activation in the output layer -- i.e., sigmoid activation only in the hidden layer (labelled ``single sigmoid'', and plotted in green).  This also leads to worse performance than the baseline configuration.

\begin{figure}
\centering
\includegraphics[width=\textwidth]{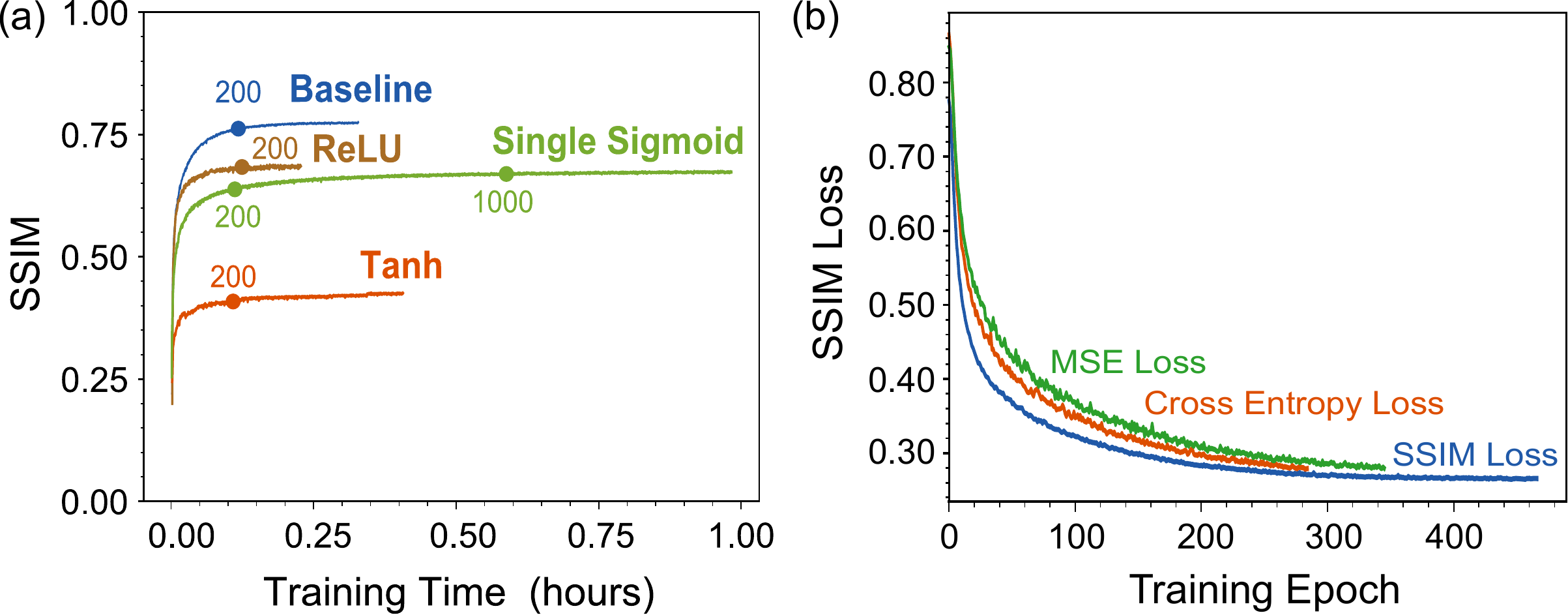}
\caption{(a) Training curves for SHL-DNNs with different kinds of activation units: double sigmoid (baseline, blue), ReLU (brown), tanh (orange), and sigmoid only in the hidden layer (green). (b) Training curves for SHL-DNNs optimized using different loss functions, evaluated in terms of SSIM loss: SSIM loss (blue), cross entropy (orange), and MSE (green).  }
\label{fig:figs3}
\end{figure}

\subsection{Choice of objective function}

The NNs in the main text are trained by maximizing the SSIM as the objective function. We also investigated other choices of objective function, including MSE loss and cross entropy loss.  These do not appear to offer any significant advantage for either type of NN.  As an example, Fig.~8(b) shows the training curves (evaluated in terms of SSIM loss) for SHL-DNNs with different choices of objective function.

\subsection{U-net filter number optimization}

In optimizing the U-Net, the performance noticeably improves as the number of filters is increased up to ``$4\times$'', (i.e., $4\times$ the number of filters used in Ref.~\cite{rahmani2018multimode}), as shown in Fig.~3(e) of the main text.  Upon increasing the filter number from $4\times$ ($31$ million trainable parameters) to $5\times$ ($49$ million trainable parameters), the saturated SSIM exhibits a very small improvement while the training time increases substantially, as shown in Fig.~9(a).  Therefore, we use $4\times$ filters in our baseline U-Net configuration.

\begin{figure}
\centering
\includegraphics[width=\textwidth]{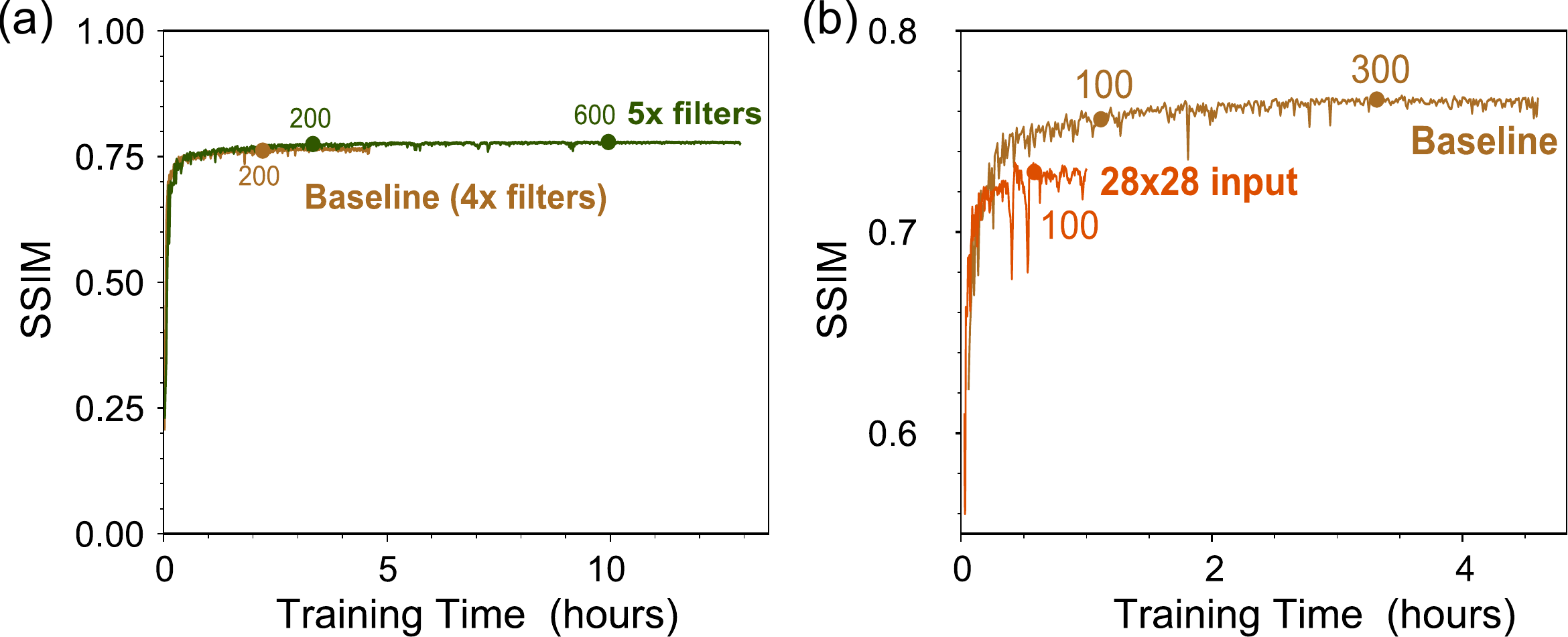}
\caption{(a) Training curve for U-Net with $4\times$ and $5\times$ the number of filters (relative Ref.~\cite{rahmani2018multimode}).  The U-Net with $4\times$ filters is the baseline configuration featured in the main text, and achieves similar performance to the U-Net with $5\times$ filters with significantly less training time.  (b) Training curves for U-Net with $64\times64$ inputs (baseline, brown) and with $28\times28$ inputs (orange).  In both cases, the ground truth and output images are $28\times28$. }
\label{fig:figs5}
\end{figure}

\subsection{U-Net with symmetric input-output}

The U-Net described in the main text is ``asymmetric'', in the sense that the input and output images have different sizes.  It is reasonable for the output images to have the same size as the ground truth images ($28\times28$), but there is no good reason to limit the input (speckle) images to that same size; they can be downscaled from the resolution of the camera image to any desired size.  In the main text, we took the input images to be $64\times64$.

We also investigated the performance of a ``symmetric'' U-Net, in which both the inputs, outputs, and ground truth images are all $28\times28$. This smaller input size necessitated removing a pair of contracting and expanding convolutional blocks (the third and fifth convolutional blocks in Fig.~1(d) in the main text). We use the same early-stopping criterion (i.e., stopping $100$ epochs after the validation curve stops improving).  As shown in Fig.~9(b), the resulting U-Net performs significantly worse, with SSIM $0.734$ compared to SSIM $0.767$ for the baseline U-Net (with 64x64 inputs).

\section{Performance of SHL-DNN and U-Net on Fashion Mnist dataset}

In Fig.~3 and 4 of the main text, we compared the SHL-DNN and U-Net performance for the MNIST digits dataset.  In Fig.~10(c), we show the same results for the MNIST-Fashion dataset.  Similar to the previous conclusions, the U-Net does not outperform the SHL-DNN in reconstructed image fidelity (as measured by the SSIM), while taking much longer to train.

\begin{figure}[h]
\centering
\includegraphics[width=\Figwt\linewidth]{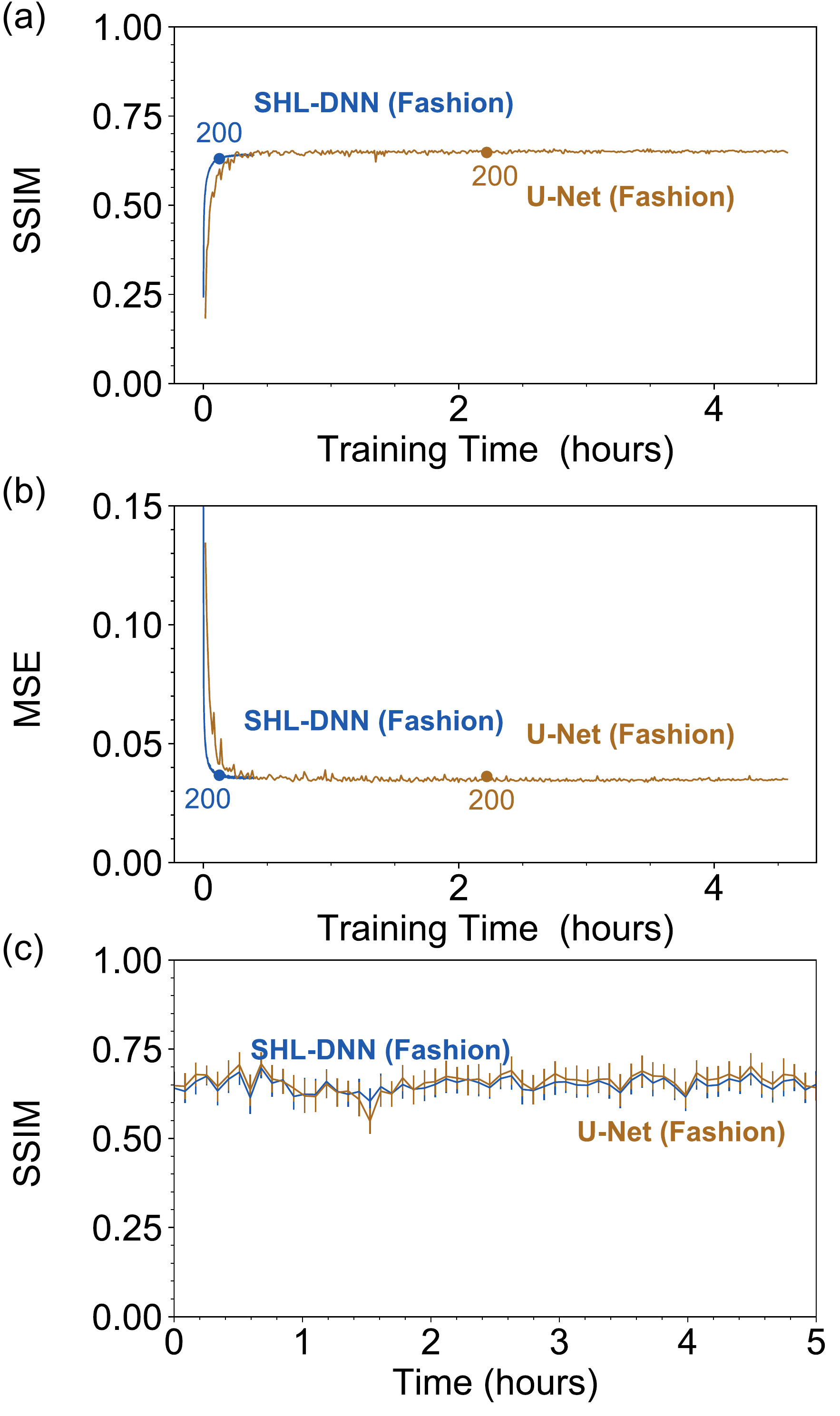}
\caption{Comparison of SHL-DNN (blue) and U-Net (brown) for the Fashion MNIST data set. (a) SSIM versus training time. (b) MSE versus training time (c) SSIM of reconstructed images using data collected up to $5$ hours subsequent to the training set.}
\label{fig:figs7}
\end{figure}

\bibliography{Ref}

\begin{thebibliography}{10}
\newcommand{\enquote}[1]{``#1''}

\bibitem{fiberendoEric}
E.~J. Seibel, R.~S. Johnston, and C.~D. Melville, \enquote{{A full-color
  scanning fiber endoscope},} in \enquote{Optical Fibers and Sensors for
  Medical Diagnostics and Treatment Applications VI,} , vol. 6083, I.~Gannot,
  ed., International Society for Optics and Photonics (SPIE, 2006), vol. 6083,
  pp. 9 -- 16.

\bibitem{fiberendoLee}
C.~Lee, C.~Engelbrecht, T.~Soper, F.~Helmchen, and E.~Seibel, \enquote{Scanning
  fiber endoscopy with highly flexible, 1 mm catheterscopes for wide-field,
  full-color imaging,} Journal of biophotonics \textbf{3}, 385--407 (2010).

\bibitem{porat2016fiberbundle}
A.~Porat, E.~R. Andresen, H.~Rigneault, D.~Oron, S.~Gigan, and O.~Katz,
  \enquote{Widefield lensless imaging through a fiber bundle via speckle
  correlations,} Opt. Express \textbf{24}, 16835--16855 (2016).

\bibitem{Shinde2017fiberbundle}
A.~Shinde, S.~M. Perinchery, and V.~M. Murukeshan, \enquote{A targeted
  illumination optical fiber probe for high resolution fluorescence imaging and
  optical switching,} Sci. Rep. \textbf{7}, 45654--45654 (2017).

\bibitem{Gover76}
A.~Gover, C.~P. Lee, and A.~Yariv, \enquote{Direct transmission of pictorial
  information in multimode optical fibers,} J. Opt. Soc. Am. \textbf{66},
  306--311 (1976).

\bibitem{choi2012scanner}
Y.~Choi, C.~Yoon, M.~Kim, T.~D. Yang, C.~Fang-Yen, R.~R. Dasari, K.~J. Lee, and
  W.~Choi, \enquote{Scanner-free and wide-field endoscopic imaging by using a
  single multimode optical fiber,} Phys. Rev. Lett. \textbf{109}, 203901
  (2012).

\bibitem{caravaca2013real}
A.~M. Caravaca-Aguirre, E.~Niv, D.~B. Conkey, and R.~Piestun,
  \enquote{Real-time resilient focusing through a bending multimode fiber,}
  Opt. Express \textbf{21}, 12881--12887 (2013).

\bibitem{gu2015design}
R.~Y. Gu, R.~N. Mahalati, and J.~M. Kahn, \enquote{Design of flexible
  multi-mode fiber endoscope,} Opt. Express \textbf{23}, 26905--26918 (2015).

\bibitem{loterie2015digital}
D.~Loterie, S.~Farahi, I.~Papadopoulos, A.~Goy, D.~Psaltis, and C.~Moser,
  \enquote{Digital confocal microscopy through a multimode fiber,} Opt. Express
  \textbf{23}, 23845--23858 (2015).

\bibitem{popoff2010measuring}
S.~Popoff, G.~Lerosey, R.~Carminati, M.~Fink, A.~Boccara, and S.~Gigan,
  \enquote{Measuring the transmission matrix in optics: an approach to the
  study and control of light propagation in disordered media,} Phys. Rev. Lett.
  \textbf{104}, 100601 (2010).

\bibitem{n2018controlled}
M.~N'Gom, N.~Estakhri, T.~B. Norris, E.~Michielssen, and R.~R. Nadakuditi,
  \enquote{Controlled transmission through highly scattering media using
  semi-definite programming as a phase retrieval computation method,} in
  \enquote{2018 Conference on Lasers and Electro-Optics (CLEO),}  (IEEE, 2018),
  pp. 1--2.

\bibitem{n2018mode}
M.~N’Gom, T.~B. Norris, E.~Michielssen, and R.~R. Nadakuditi, \enquote{Mode
  control in a multimode fiber through acquiring its transmission matrix from a
  reference-less optical system,} Opt. Lett. \textbf{43}, 419--422 (2018).

\bibitem{aisawa1991remote}
S.~Aisawa, K.~Noguchi, and T.~Matsumoto, \enquote{Remote image classification
  through multimode optical fiber using a neural network,} Opt. Lett.
  \textbf{16}, 645--647 (1991).

\bibitem{matsumoto1990proposal}
T.~Matsumoto, M.~Koga, K.~Noguchi, and S.~Aizawa, \enquote{Proposal for
  neural-network applications to fiber-optic transmission,} in \enquote{1990
  IJCNN International Joint Conference on Neural Networks,}  (IEEE, 1990), pp.
  75--80.

\bibitem{marusarz2001neural}
R.~K. Marusarz and M.~R. Sayeh, \enquote{Neural network-based multimode
  fiber-optic information transmission,} Applied optics \textbf{40}, 219--227
  (2001).

\bibitem{borhani2018learning}
N.~Borhani, E.~Kakkava, C.~Moser, and D.~Psaltis, \enquote{Learning to see
  through multimode fibers,} Optica \textbf{5}, 960--966 (2018).

\bibitem{rahmani2018multimode}
B.~Rahmani, D.~Loterie, G.~Konstantinou, D.~Psaltis, and C.~Moser,
  \enquote{Multimode optical fiber transmission with a deep learning network,}
  Light: Sci. \& Appl. \textbf{7}, 1--11 (2018).

\bibitem{fan2019deep}
P.~Fan, T.~Zhao, and L.~Su, \enquote{Deep learning the high variability and
  randomness inside multimode fibers,} Opt. Express \textbf{27}, 20241--20258
  (2019).

\bibitem{yang2019deep}
M.~Yang, Z.-H. Liu, Z.-D. Cheng, J.-S. Xu, C.-F. Li, and G.-C. Guo,
  \enquote{Deep hybrid scattering image learning,} J. Phys. D: Appl. Phys.
  \textbf{52}, 115105 (2019).

\bibitem{rahmani2019multimode}
B.~Rahmani, D.~Loterie, E.~Kakkava, N.~Borhani, U.~Te{\u{g}}in, D.~Psaltis, and
  C.~Moser, \enquote{Multimode fiber projector,}  (2019). Arxiv 1907.00126.

\bibitem{Kakkava2019}
E.~Kakkava, B.~Rahmani, N.~Borhani, U.~Te{\u{g}}in, D.~Loterie,
  G.~Konstantinou, C.~Moser, and D.~Psaltis, \enquote{Imaging through multimode
  fibers using deep learning: The effects of intensity versus holographic
  recording of the speckle pattern,} Opt. Fiber Tech. \textbf{52}, 101985
  (2019).

\bibitem{braspenning1995artificial}
P.~J. Braspenning, F.~Thuijsman, and A.~J. M.~M. Weijters, \emph{Artificial
  neural networks: an introduction to ANN theory and practice}, vol. 931
  (Springer Science \& Business Media, 1995).

\bibitem{Goodfellow-et-al-2016}
I.~Goodfellow, Y.~Bengio, and A.~Courville, \emph{Deep Learning} (MIT Press,
  2016).

\bibitem{Rawat2017}
W.~Rawat and Z.~Wang, \enquote{Deep convolutional neural networks for image
  classification: A comprehensive review,} Neural Computation \textbf{29},
  2352--2449 (2017).

\bibitem{li2018deep}
Y.~Li, Y.~Xue, and L.~Tian, \enquote{Deep speckle correlation: a deep learning
  approach toward scalable imaging through scattering media,} Optica
  \textbf{5}, 1181--1190 (2018).

\bibitem{li2018imaging}
S.~Li, M.~Deng, J.~Lee, A.~Sinha, and G.~Barbastathis, \enquote{Imaging through
  glass diffusers using densely connected convolutional networks,} Optica
  \textbf{5}, 803--813 (2018).

\bibitem{Turpin2018Control}
A.~Turpin, I.~Vishniakou, and J.~d.~Seelig, \enquote{Light scattering control
  in transmission and reflection with neural networks,} Opt. Express
  \textbf{26}, 30911--30929 (2018).

\bibitem{sun2019image}
Y.~Sun, J.~Shi, L.~Sun, J.~Fan, and G.~Zeng, \enquote{Image reconstruction
  through dynamic scattering media based on deep learning,} Opt. express
  \textbf{27}, 16032--16046 (2019).

\bibitem{Redding2012}
B.~Redding and H.~Cao, \enquote{Using a multimode fiber as a high-resolution,
  low-loss spectrometer,} Opt. Lett. \textbf{37}, 3384--3386 (2012).

\bibitem{Pauwels2019}
J.~Pauwels, G.~Van~der Sande, and G.~Verschaffelt, \enquote{Space division
  multiplexing in standard multi-mode optical fibers based on speckle pattern
  classification,} Sci. Rep. \textbf{9}, 17597 (2019).

\bibitem{Linsley2018}
D.~Linsley, J.~Kim, V.~Veerabadran, C.~Windolf, and T.~Serre, \enquote{Learning
  long-range spatial dependencies with horizontal gated recurrent units,} in
  \enquote{32nd Conference on Neural Information Processing Systems (NeurIPS
  2018),}  (ACM, 2018), p. 152.

\bibitem{ronneberger2015u}
O.~Ronneberger, P.~Fischer, and T.~Brox, \enquote{U-net: Convolutional networks
  for biomedical image segmentation,} in \enquote{International Conference on
  Medical image computing and computer-assisted intervention,}  (Springer,
  2015), pp. 234--241.

\bibitem{ZhangGAN}
X.~Zhang, Z.~Yu, Z.~Meng, K.~Ding, Z.~Ju, and K.~Xu, \enquote{Experimental
  demonstration of a multimode fiber imaging system based on generative
  adversarial networks,} in \enquote{Asia Communications and Photonics
  Conference (ACPC) 2019,}  (Optical Society of America, 2019), p. T4A.4.

\bibitem{wang2004image}
Z.~Wang, A.~C. Bovik, H.~R. Sheikh, and E.~P. Simoncelli, \enquote{Image
  quality assessment: from error visibility to structural similarity,} IEEE
  Trans. Image Processing \textbf{13}, 600--612 (2004).

\bibitem{LanczosResampling}
B.~N. Madhukar and R.~Narendra, \enquote{Lanczos resampling for the digital
  processing of remotely sensed images,} in \enquote{Proceedings of
  International Conference on VLSI, Communication, Advanced Devices, Signals
  {\&} Systems and Networking (VCASAN-2013),} , V.~S. Chakravarthi, Y.~J.~M.
  Shirur, and R.~Prasad, eds. (Springer India, India, 2013), pp. 403--411.

\bibitem{lecun2010mnist}
Y.~LeCun, C.~Cortes, and C.~Burges, \enquote{Mnist handwritten digit database,}
  ATT Labs [Online]. Available: http://yann.lecun.com/exdb/mnist \textbf{2}
  (2010).

\bibitem{xiao2017fashionmnist}
H.~Xiao, K.~Rasul, and R.~Vollgraf, \enquote{Fashion-mnist: a novel image
  dataset for benchmarking machine learning algorithms,}  (2017). Arxiv
  1708.07747.

\bibitem{keras}
F.~Chollet \emph{et~al.}, \enquote{Keras,}
  \url{https://github.com/fchollet/keras} (2015).

\bibitem{simonyan2014deep}
K.~Simonyan and A.~Zisserman, \enquote{Very deep convolutional networks for
  large-scale image recognition,}  (2014). Arxiv 1409.1556.

\bibitem{caramazza2019transmission}
P.~Caramazza, O.~Moran, R.~Murray-Smith, and D.~Faccio, \enquote{Transmission
  of natural scene images through a multimode fibre,} Nat. Commun. \textbf{10},
  1--6 (2019).

\end{thebibliography}
\end{document}